\documentclass[graybox]{svmult}

\usepackage{type1cm}        
\usepackage{makeidx}         
\usepackage{graphicx}        
\usepackage{multicol}        
\usepackage[bottom]{footmisc}

\usepackage{newtxtext}       %
\usepackage{newtxmath}       

\usepackage{placeins}
\usepackage{hyperref}
\usepackage{subfig}
\usepackage{booktabs}
\usepackage{multirow}
\usepackage{cancel}
\usepackage{moresize}

\usepackage[normalem]{ulem}
\usepackage{tikz}
\usetikzlibrary{arrows.meta}
\tikzset{%
   >={Latex[width=2mm,length=2mm]},
 base/.style = {rectangle, rounded corners, draw=black, minimum width=2.2cm, text width=4cm},
 inputOutput/.style = {base, dotted, fill=white, draw=black, text centered, node distance=0.85cm},
 generator/.style = {base, fill=pink, text centered, node distance=0.95cm},
 discriminator/.style = {base, fill=cyan, text centered, node distance=1.15cm},
 legend/.style = {base, dashed, fill=white},    
 classifier/.style = {base, fill=olive, text centered, node distance=1.29cm}
 }

\def\BibTeX{{\rm B\kern-.05em{\sc i\kern-.025em b}\kern-.08em
    T\kern-.1667em\lower.7ex\hbox{E}\kern-.125emX}}

\makeindex             

\begin{document}

\title*{Evaluating the Quality and Diversity of DCGAN-based Generatively Synthesized Diabetic Retinopathy Imagery}
\titlerunning{Evaluating DCGAN-based Synthetic PDR Images}
\author{Cristina-Madalina Dragan, Muhammad Muneeb Saad, Mubashir Husain Rehmani, and Ruairi O'Reilly}
\authorrunning{Cristina-Madalina Dragan et al.}
\institute{Cristina-Madalina Dragan was with Munster Technological University, Cork, Ireland. \email{cristina.madalina.dragan@gmail.com}
\and \newline Muhammad Muneeb Saad, Mubashir Husain Rehmani, and Ruairi O'Reilly are with Munster Technological University, Cork, Ireland. \email{muhammad.saad@mycit.ie, mubashir.rehmani@mtu.ie, ruairi.oreilly@mtu.ie.}}
%
%
\maketitle

\abstract{\emph{Publicly available diabetic retinopathy (DR) datasets are imbalanced, containing limited numbers of images with DR. This imbalance contributes to overfitting when training machine learning classifiers. The impact of this imbalance is exacerbated as the severity of the DR stage increases, affecting the classifiers' diagnostic capacity. The imbalance can be addressed using Generative Adversarial Networks (GANs) to augment the datasets with synthetic images. Generating synthetic images is advantageous if high-quality and diverse images are produced. To evaluate the quality and diversity of synthetic images, several evaluation metrics, such as Multi-Scale Structural Similarity Index (MS-SSIM), Cosine Distance (CD), and Fréchet Inception Distance (FID), are used. Understanding the effectiveness of each metric in evaluating the quality and diversity of synthetic images is critical to select images for augmentation. To date, there has been limited analysis of the appropriateness of these metrics in the context of biomedical imagery. This work contributes an empirical assessment of these evaluation metrics as applied to synthetic Proliferative DR imagery generated by a Deep Convolutional GAN (DCGAN). Furthermore, the metrics' capacity to indicate the quality and diversity of synthetic images and their correlation with classifier performance are examined. This enables a quantitative selection of synthetic imagery and an informed augmentation strategy, which are often lacking in the literature. Results indicate that FID is suitable for evaluating the quality, while MS-SSIM and CD are suitable for evaluating the diversity of synthetic imagery. Furthermore, the superior performance of Convolutional Neural Network (CNN) and EfficientNet classifiers, as indicated by the $F_1$ and AUC scores, for the augmented datasets compared to the original dataset demonstrate the efficacy of synthetic imagery to augment the imbalanced dataset while improving the classification scores.}}

\section{Introduction}

\begin{figure}[hbt!]
\centering
    \subfloat[]
         {\includegraphics[width=2cm]{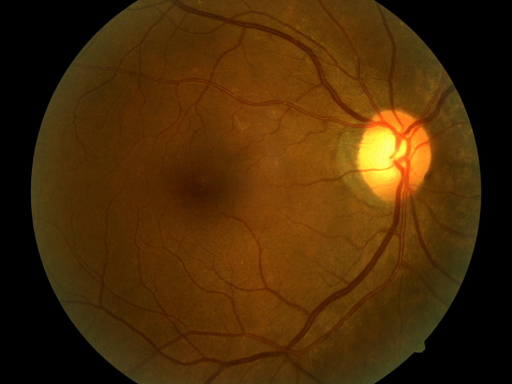}}
     \hspace{0.1cm}
    \subfloat[]
         {\includegraphics[width=2.25cm]{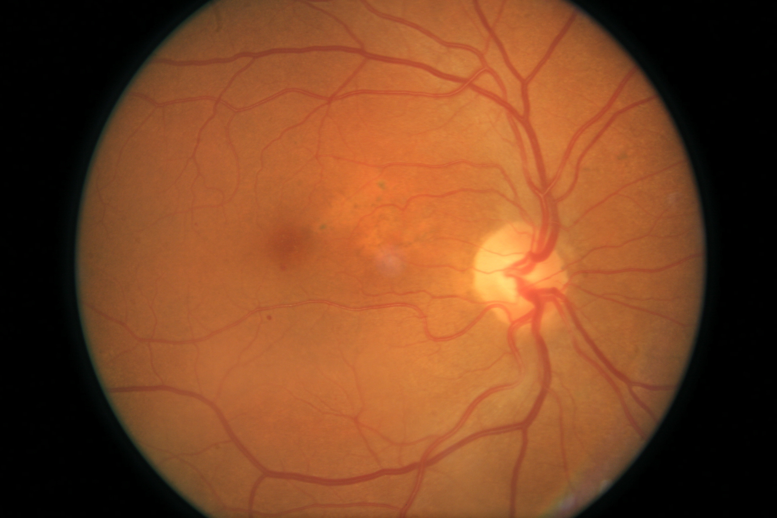}}
     \hspace{0.1cm}
     \subfloat[]
         {\includegraphics[width=2.25cm]{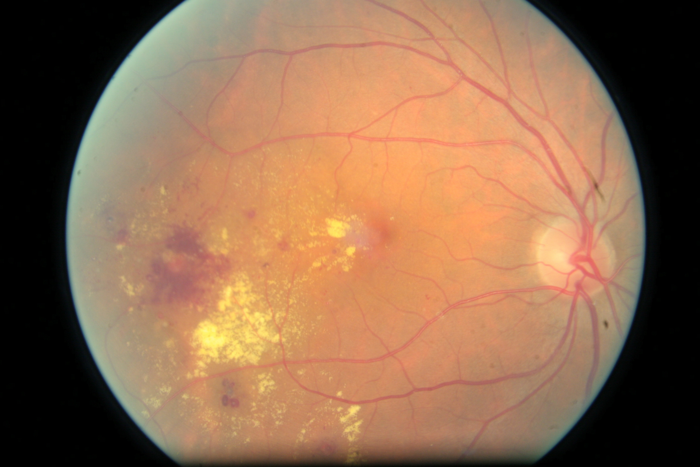}}
     \hspace{0.1cm}
     \subfloat[]
         {\includegraphics[width=2cm]{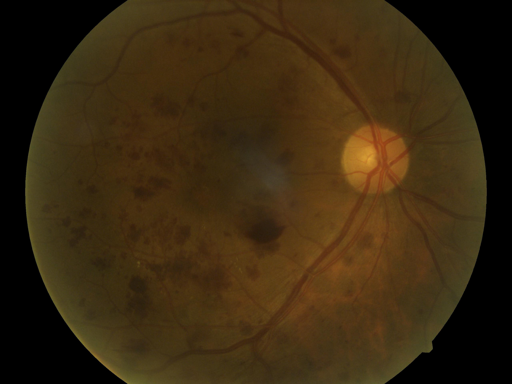}}
     \hspace{0.1cm}
     \subfloat[]
         {\includegraphics[width=2cm]{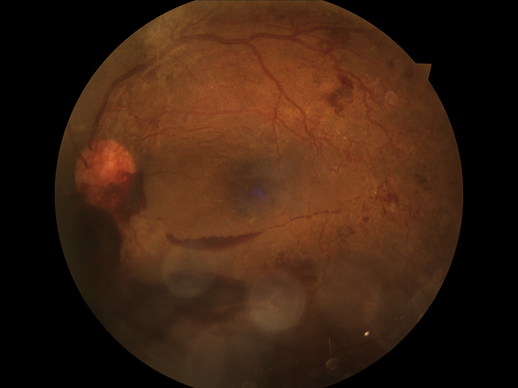}}
    \caption{Retinal fundus images with different stages of DR: (a) No DR (b) Mild NPDR (c) Moderate NPDR (d) Severe NPDR (e) PDR}
    \label{fig:drStages}
\end{figure}

Diabetic retinopathy (DR) is a complication caused by high blood sugar levels over a prolonged period that is estimated to affect 415 million people globally \cite{Cavan_2017} and can lead to blindness if it is not treated timely. The diagnosis of DR is made based on the analysis of retinal fundus imagery, where lesions specific to DR are identified. The severity of DR is evaluated using the international severity grading scale (ISGR) \cite{Wilkinson_2003}, which has four stages: Mild Non-Proliferative DR (Mild NPDR), Moderate NPDR, Severe NPDR, and Proliferative DR (PDR) (see Fig.~\ref{fig:drStages}). The increasing prevalence of the disease and the lack of medical personnel capable of diagnosing it highlight the need for computer-aided diagnostics to assist healthcare professionals \cite{Arora_2019, Ghosh_2017, Ni_2019}. 

Artificial intelligence (AI) techniques have become important in finding solutions to modern engineering problems. AI has been utilized in the domain of biomedical imagery for disease analysis and the interpretation of clinical data \cite{ali2022combating}. Healthcare has become increasingly dependent on computer-aided diagnosis (CAD), a computer-based application that assists clinicians \cite{chen2022generative}. There is a vast contribution made by AI-based classifiers, including Support Vector Machine (SVM), Logistic Regression (LR), Artificial Neural Networks (ANNs), and deep learning models such as Convolutional Neural Networks (CNNs), to assist clinicians through automated analysis of numerous diseases such as diabetes, cancer, and COVID-19 using biomedical imagery \cite{Arora_2019, Ghosh_2017, Ni_2019, saad2022addressing}. Deep learning models can provide more effective disease analysis than alternate techniques. However, these models require large quantities of training data to enable effective classification, which is a challenging problem in the domain of biomedical imagery \cite{Rahman_2013, Shorten_2019}.

Publicly available retinal fundus image datasets are imbalanced, containing significantly more images without DR than images with DR, as indicated in Table \ref{tab:datasets}. The availability of images decreases as the severity of DR increases, with severe NDPR and PDR accounting for a total of 6\% of the available images. Imbalanced datasets contain images with skewed classes. The skewness in imbalanced datasets refers to the asymmetry distribution of images across different classes \cite{Shorten_2019}. Therefore, when the severity of DR disease increases then this imbalance causes overfitting of the classifiers to the class representing the most severe stage of the disease \cite{saini2020deep}. The Kaggle~\cite{Cuadros_2009} dataset (See table \ref{tab:datasets}) is highly imbalanced and has the largest number of images as compared to the other datasets. This dataset is important as it contains images from patients of different ethnicities. 

In the domain of biomedical imagery, data imbalance is a challenging problem as it deals with imagery that contains salient features indicative of diseases that directly impact human lives \cite{chen2022generative}. As such, addressing the issue of data imbalance is considered a worthwhile endeavor. One potential solution is to augment training data with synthetic imagery belonging to the underrepresented classes \cite{oza2022image}.

\begin{table}[hbt!]
    \centering
    \caption{\textbf{Publicly available datasets containing retinal fundus images at different stages of DR.}}
    \begin{tabular}{p{2.4cm}p{1.3cm}p{1cm}p{1.5cm}p{2.2cm}p{1.7cm}p{0.8cm}} 
    \toprule
     \textbf{Dataset} & \textbf{No. Image} & \textbf{No DR} & \textbf{Mild NPDR} & \textbf{Moderate NPDR} & \textbf{Severe NPDR} & \textbf{PDR} \\
    \midrule
       Kaggle~\cite{Cuadros_2009} & 
            88702   & 
            65343   &
            6205    &
            13153   &
            2087    &
            1914\\ 
       APTOS~\cite{Kaggle_Aptos} & 
            3662    & 
            1805    & 
            370     & 
            999     & 
            193     & 
            295\\
       FGADR~\cite{Zhou_2021} & 
            2842    & 
            244     & 
            337     & 
            1161    & 
            752     & 
            348\\
       Messidor-2 ~\cite{Decenciere_2014,  Abramoff_2013} & 
            1744    & 
            1017    & 
            270     & 
            347     &
            75      & 
            35 \\
       Messidor$^{\mathrm{a}}$~\cite{Decenciere_2014} & 
            1200    & 
            546     & 
            153     & 
            247     & 
            254     &
            - \\
       IDRID~\cite{Porwal_2018} & 
            516     & 
            168     & 
            25      & 
            168     & 
            93      &
            62 \\
       DR2~\cite{Pires_2014} & 
            435 &
            98  &
            337 &
            -   &
            -   &
            - \\
       1000 Fundus~\cite{Cen_2021} &
           144  &
           38   &
           18   &
           49   &
           39   &
           - \\
       STARE~\cite{Hoover_2000} &
            113 &
            41  &
            51  &
            21  &
            -   &
            - \\
       HRFID~\cite{Odstrcilik_2013} & 
            30  &
            15  &
            15  &
            -   &
            -   &
            - \\
        \midrule
        \textbf{Total$^{\mathrm{a}}$} & 
             98188	&
             68769 &
             7628 &
             15898 &
             3239 &
             2654 \\
            As \% &
                    &
             70 &
             7.8 &
             16.2 &
             3.3 &
             2.7 \\
        \bottomrule
        \multicolumn{7}{l}{DR: Diabetic Retinopathy; NPDR: Non-Proliferative DR; PDR: Proliferative DR} \\
        \multicolumn{7}{l}{\footnotesize{$^{\mathrm{a}}$ Messidor excluded from total as 1058 images overlap with the Messidor-2 dataset}}
    \end{tabular}
    
\label{tab:datasets}
\end{table}

Synthetic biomedical imagery can be derived from generative models. Several generative models such as variational autoencoders \cite{sengupta2020funsyn}, diffusion models \cite{shi2023dissolving}, and Generative Adversarial Networks (GANs) \cite{Zhou_2020} have been utilized to generate retinal fundus synthetic images. For generating synthetic images, autoencoders produce blurry images while diffusion models are trained slowly with high-computational cost as compared to GANs \cite{kebaili2023deep}. GANs are generative models consisting of two neural networks, a generator, and a discriminator. The generator aims to produce realistic synthetic images, and the discriminator's aim is to distinguish between real and synthetic images. 

The GAN-based generated synthetic images are evaluated using two critical criteria, the quality of the images, indicating how representative they are of real images of that class, and the diversity of the images, indicating how broad and uniform the coverage of their feature distribution is compared to the real images. The quality of synthetic imagery is characterized by an alignment of its feature distribution with the class label \cite{Zhou_2020, Costa_2018, Yu_2019} and a low level of noise, blurriness, and distortions \cite{Thung_2009}. The diversity of synthetic images is characterized by the level of similarity to each other \cite{Shmelkov_2018}.

The quality and diversity of synthetic images have a significant impact on the performance of classifiers when these images are used for augmenting limited and imbalanced datasets. A classifier will not learn the features representative of different classes if the training dataset is augmented with low-quality synthetic images. Similarly, the classifier will incorrectly classify images containing feature distributions that belong to the less represented areas if the training dataset is augmented with less diversified synthetic images. When synthetic images are used in training a classifier, it is essential that they are of high quality and that the features representative of a class are sufficiently diverse.
 
The quality of synthetic imagery represents the level of similarity to the real imagery and the diversity represents the level of dissimilarity between the synthetic imagery. There are several metrics used for evaluating the similarity between images such as peak signal-to-noise ratio (PSNR) \cite{Borji_2019}, structural similarity index (SSIM) \cite{Wang_2003} \cite{Borji_2019}, multi-scale structural similarity index (MS-SSIM) \cite{Wang_2003} \cite{odena2017conditional}, cosine distance (CD) \cite{salimans2018improving}, and Fréchet inception distance (FID) \cite{Heusel_2017}.

In the literature, these metrics are categorized based on qualitative and quantitative measures \cite{Borji_2019}. Qualitative measures require subjective information such as visual examination of synthetic images by humans which is time-consuming and cumbersome \cite{Borji_2019}. It includes evaluation metrics such as nearest neighbors \cite{Goodfellow_2014}, rapid scene categorization \cite{denton2015deep}, and preference judgment \cite{snell2017learning}. On the other hand, quantitative measures do not require subjective information and only rely on objective information such as the diversity and quality of images, when evaluating synthetic images using quantitative evaluation metrics \cite{Borji_2019}.

In this work, quantitative metrics such as MS-SSIM, CD, and FID are used. A combination of these metrics provides quantitative measures using perceptual features (MS-SSIM) and the distance between image pixels (CD and FID) to evaluate the quality and diversity of synthetic images.

\begin{figure}[htp!]
    \centering
    \includegraphics[width=1\textwidth]{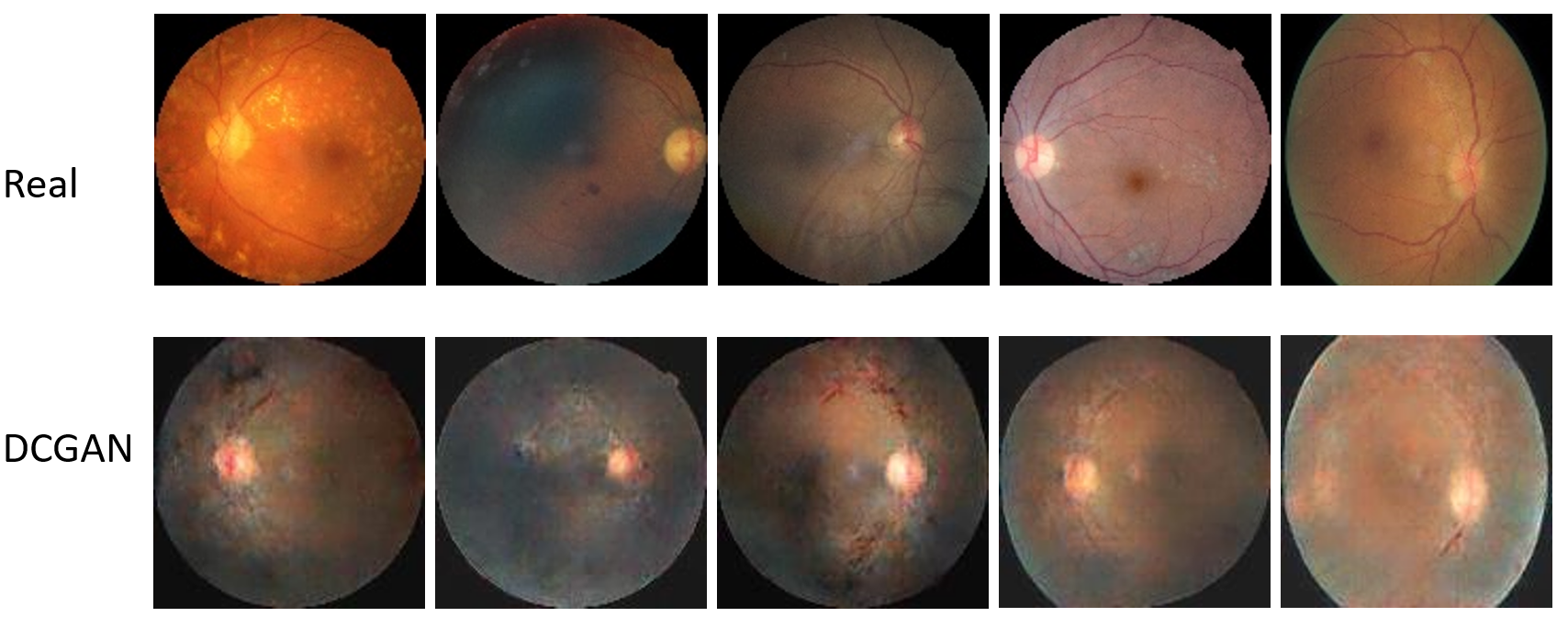}
    \caption{Real and synthetically generated PDR image samples of Deep Convolutional GAN depicting lesions of different size, shape, and location.}
    \label{Fig.real_synthetic}
\end{figure}

PDR is the most severe stage of DR with the lowest quantity of publicly available imagery, as such this work narrows its focus to the generation of synthetic PDR imagery. PDR is a serious eye complication of diabetes that can lead to severe vision loss or even blindness. It occurs when abnormal blood vessels grow in the retina, the light-sensitive tissue at the back of the eye, in response to high blood sugar levels. It is important for people with diabetes to have regular eye exams to detect PDR at early stages and prevent vision loss \cite{Cavan_2017}. In the context of PDR, a high-quality synthetic image depicts the presence of valid lesions specific to PDR, and a high level of diversity is indicated by the presence of lesions of different sizes, shapes, and locations, as shown in Fig. \ref{Fig.real_synthetic}.

There are a few evaluation metrics, such as SSIM and PSNR, that evaluate the quality of generated images by comparing them to ground truth images using image pixel values. Other evaluation metrics, such as SWD and FID, quantify the distance between images to evaluate the quality of images. However, it is important to evaluate the diversity of images because GANs should generate synthetic images that are as diverse as real images. Therefore, evaluation metrics that can quantify the quality and diversity of synthetic images should be used. Several methods, such as classification scores achieved through neural networks and the analysis of radiologists, for evaluating synthetic images have been adopted, as indicated in Table \ref{tab:GansMedicalImages}. It is challenging to find evaluation metrics that explicitly evaluate the quality and diversity of synthetic images using significant image features by comparing synthetic images to real images.

Technical discussions on the selection and suitability of metrics for evaluating the quality and diversity of GAN-based synthetic retinal fundus images are lacking in the literature. A limiting factor in existing works is that the method of selecting the synthetic images used for data augmentation is not specified. In the works indicated in Table \ref{tab:GansMedicalImages}, it is not assessed whether the metrics used are suitable to evaluate the quality and diversity of the synthetic imagery. 

In this work, an empirical assessment of MS-SSIM, CD, and FID metrics is conducted to assess the suitability of these metrics in evaluating the quality and diversity of synthetic DR images generated by the DCGAN. Moreover, this work contributes to: (i) a critical analysis of quantitative evaluation metrics’ capacity to identify if imagery contains features corresponding to its class label; (ii) an investigation of DCGAN’s capacity to generate diversified and high-quality synthetic PDR images; (iii) an assessment of DCGAN's synthetic images to improve the classification performance of classifiers such as CNN and EfficientNet; and (iv) an evaluation of the relationship between diversity and quality of the synthetic imagery as indicated by MS-SSIM, CD, FID, and classification performance.

It is envisaged that understanding which evaluation metrics are suitable for evaluating the quality and diversity of synthetic retinal fundus imagery will enable an improved selection of synthetic imagery to augment the training dataset of a classifier.

\begin{table}[htp!]
\centering
\caption{\textbf{Generation of biomedical imagery utilizing GANs. The classification performance gain is noted where generated images were used in training a classifier.}
}

\begin{tabular}{p{1.3cm}p{1cm}p{1.2cm}p{1.1cm}p{0.75cm}p{1.1cm}p{2cm}p{1.4cm}p{0.2cm}}

\toprule

\textbf{Medical Img.} & \textbf{Ref}~\textsubscript{\textbf{Year}} & \textbf{GANs} & \textbf{Dataset} & \textbf{Img. Res.} & \textbf{Clf.} & \textbf{Clf. (Gain)} & \textbf{Quality} & \textbf{Div.} \\

\midrule

\multirow{2}{*}{\multirow{2}{2cm}{DR Retinal \\ Fundus}} &  
\cite{Zhou_2020}~\textsubscript{2020}
    & CGAN
    & Kaggle,
    & \multirow{2}{2cm}{1280x\\1280}
    & VGG-16, ResNet-50, I-v3
    & \multirow{2}{2.9cm}{Acc:(0.01)\\ Kappa:\\(0.01-0.02)} & \multirow{4}{2cm}{FID, SWD, \\ visualized \\ by Ophthal-\\ mologists} & N/A \\
&&&&&& \\
& \cite{Balasubramanian_2020}~\textsubscript{2020}
    & DCGAN
    & Kaggle
    & \multirow{2}{2cm}{128x\\128}
    & CNN 
    & \multirow{3}{2.9cm}{Mac-aver \\ Prec:(-0.01-0) \\ Mac-aver \\ Rec: (0.01) \\ Mac-aver \\ F1:(0-0.01)} & CD & CD \\
&&&&&& \\
&&&&&& \\
&&&&&& \\
&&&&&& \\
&&&&&& \\
&&&&&& \\
&&&&&& \\
\multirow{6}{*}{\multirow{2}{2cm}{Retinal \\ Fundus}} &
\cite{Lim_2020}~\textsubscript{2020}
    & StyleGAN
    & Kaggle
    & \multirow{2}{2cm}{512x\\512}
    & ResNet-50
    & N/A & N/A & N/A \\
&&&&&& \\
& \cite{Burlina_2019}~\textsubscript{2019}
    & ProGAN
    & AREDS
    & \multirow{2}{2cm}{512x\\512}
    & ResNet-50 
    & N/A & \multirow{3}{2.9cm}{Visualized \\ by Eye\\Specialists} & N/A \\
&&&&&& \\
&&&&&& \\
& \cite{HaoQi_2020}~\textsubscript{2020}
    & CGAN 
    & DRIVE \cite{Staal_2004}
    & \multirow{2}{2cm}{512x\\512}
    & N/A 
    & N/A & N/A & N/A \\
&&&&&& \\
& \cite{Costa_2018}~\textsubscript{2018}
    & CGAN
    & Messidor
    & \multirow{2}{2cm}{256x\\256}
    & N/A & N/A & ISC score & N/A \\
&&&&&& \\
& \cite{Yu_2019}~\textsubscript{2019}
    & Pix2pix, Cycle-GAN
    & \multirow{3}{4cm}{DRISHTI \\ -GS\cite{Sivaswamy_2015}, \\ DRIVE}
    & \multirow{4}{2cm}{512x\\512, \\256x\\256}
    & N/A 
    & N/A & SSIM, PSNR & N/A \\
&&&&&& \\
& \cite{Diaz-Pinto_2019}~\textsubscript{2019}
    & SS-DCGAN 
    & Irrelevant
    & \multirow{2}{2cm}{128x\\128}
    & SS-DCGAN 
    & N/A & \multirow{2}{2.9cm}{LSE, \\ T-SNE} & AVP, MSE \\
&&&&&& \\
&&&&&& \\
&&&&&& \\
DR lesions & \cite{Chen_2019}~\textsubscript{2019} 
    & DCGAN 
    & Irrelevant
    & 32x32
    & CNN 
    & \multirow{2}{2cm}{Sens:(0.05-0.2) \\ Spec:(0.07-0.23)} & N/A & N/A \\
&&&&&& \\
&&&&&& \\
Brain PET & \cite{Islam_2020}~\textsubscript{2020}
    & DCGAN 
    & Irrelevant
    & \multirow{2}{2cm}{128×\\128}
    & CNN 
    & Acc:(0.1) & \multirow{2}{2.9cm}{PSNR, \\ SSIM} & N/A \\
&&&&&& \\
&&&&&& \\
Liver CT & \cite{Frid_2018}~\textsubscript{2018}
    & DCGAN, AC-GAN 
    & Irrelevant
    & 64x64
    & CNN 
    & \multirow{2}{2cm}{Sens:(0.02-0.13) \\ Spec:(-0.01-0.07)} & Visualized by Radio. & N/A \\
&&&&&\\
Brain MR & \cite{Han_2019}~\textsubscript{2019}
    & Cond. PGGAN
    & Irrelevant
    & \multirow{2}{2cm}{256×\\256}
    & YOLOv3 CNN \cite{Redmon_2018}  
    & Sens: (0.01 - 0.1) & VTT & T-SNE \\
&&&&&& \\
& \cite{Han_Rundo_2019}~\textsubscript{2019}
    & PGGAN, SimGAN
    & Irrelevant
    & \multirow{2}{2cm}{224x\\224}
    & ResNet-50  
    & \multirow{3}{2cm}{Acc:(-0.08-0.1) \\ Sens:(-0.08-0.08) Spec:\\(-0.11-0.14)}& \multirow{2}{2cm}{VTT, \\ T-SNE} & T-SNE \\
&&&&&& \\
&&&&&& \\
\bottomrule
\multicolumn{9}{l}{Acc: Accuracy; Clf: Classifier; Cond: Conditional; Div: Diversity; DR: Diabetic Retinopathy} \\
\multicolumn{9}{l}{Img: Image; Mac-aver: Macro-averaged; Prec: Presion; Ref: Reference; Res: Resolution} \\
\multicolumn{9}{l}{Rec: Recall; Radio: Radiologists; Sens: Sensitivity; Spec: Specificity; VTT: Visual Turing Test} \\
\end{tabular}
\label{tab:GansMedicalImages}
\end{table}

\section{Related Work}

Several state-of-the-art works that use CNNs for automatically classifying DR via retinal fundus imagery are denoted in Table \ref{tab:classifiers}. In order for CNNs to generalize across DR stages whilst achieving a performant classification accuracy, these classifiers need to be trained on a large and balanced dataset~\cite{Pei_2020, Kotsiantis_2006}.

In detailing the CNNs in Table \ref{tab:classifiers} macro-averaged F1 score (mac. F1) was used to compare their performance, as all classes are treated equally \cite{Sokolova_2009}. This is particularly important due to the data imbalance and the minority class (PDR) being the most severe stage of DR. The macro-averaged F1 score was calculated based on the mean of the F1 scores for every class \cite{Zhongze_2021}.

\begin{table}[hbt!]
\centering
\caption{\textbf{CNN-based classification of DR stages based on the ISGR.}}
\begin{tabular}{p{1.2cm}p{2.5cm}p{1.1cm}p{1.1cm}p{1.1cm}p{2.1cm}p{1.4cm}} 
\toprule
\textbf{Ref~\textsubscript{Year}} & \textbf{Model} & \textbf{Dataset} & \textbf{Acc.} & \textbf{Mac. F1} & \textbf{Data Augment.} & \textbf{Image Res.} \\
\midrule
\cite{Gayathri_2020}~\textsubscript{2020} 
    & CNN and DT 
    & Kaggle 
    & 99.99 
    & 0.999$^{\mathrm{a}}$ 
    & - 
    & 227x227 \\
\cite{Sayres_2019}~\textsubscript{2019} 
    & Inception V4 & Private 
    & 88.4 
    & 0.678$^{\mathrm{a}}$ 
    & - 
    & 779x779 \\
\cite{Zeng_2019}~\textsubscript{2019} 
    & Siamese CNN 
    & Kaggle 
    & 84.25$^{\mathrm{a}}$ 
    & 0.603$^{\mathrm{a}}$ 
    & Trad. 
    & 299x299 \\
\cite{Ghosh_2017}~\textsubscript{2017} 
    & CNN/Denoising & Kaggle 
    & 85 
    & 0.566$^{\mathrm{a}}$ 
    & Trad. 
    & 512x512 \\
\cite{Qummar_2019}~\textsubscript{2019} 
    & Ensemble/TL 
    & Kaggle & 80.8 
    & 0.532 
    & Trad. 
    & 512x512 \\
\cite{Kwasigroch_2018}~\textsubscript{2018} 
    &  Deep CNN 
    & Kaggle 
    & 50.8 
    & 0.482$^{\mathrm{a}}$ 
    & Trad. 
    & 224x224 \\
\cite{Pratt_2016}~\textsubscript{2016} 
    & Deep CNN & Kaggle 
    & 73.76$^{\mathrm{a}}$ 
    & 0.335$^{\mathrm{a}}$ 
    & Trad. (+CW) 
    & 512x512 \\
\cite{Balasubramanian_2020}~\textsubscript{2020} 
    & Deep CNN 
    & Kaggle
    & 0.693$^{\mathrm{a}}$ 
    & 0.268$^{\mathrm{a}}$ 
    & GANs 
    & 128x128\\
\cite{Arora_2019}~\textsubscript{2019} 
    & CNN & Kaggle 
    & 74 
    & - 
    & Trad. 
    & 128x128 \\
\cite{Ni_2019}~\textsubscript{2019} 
    & Deep CNN
    & Kaggle 
    & 87.2 
    & - 
    & Sampling and IW. 
    & 600x600 \\
\cite{Zhou_2020}~\textsubscript{2020} 
    & VGG-16, ResNet-50, I-v3,
AFN, Zoom-in 
    & Kaggle, FGARD 
    & 82.45-89.16
    & - 
    & GANs 
    & 1280x1280 \\
\bottomrule
\multicolumn{7}{l}{Augment: Augmentation; Acc: Accuracy; CW: Class Weights; DT: Decision Trees} \\
\multicolumn{7}{l}{IW: Instance Weights; Mac. F1: Macro F1; Ref: Reference; Res: Resolution; Trad: Traditional} \\
\multicolumn{7}{l}{TL: Transfer Learning; \footnotesize{$^{\mathrm{a}}$ denotes the derived performance from confusion matrix}}
\end{tabular}
\label{tab:classifiers}
\end{table}

Data augmentation is one approach to address data imbalance when training a CNN. It consists of adding synthetic or modified versions of the original images from the underrepresented classes to the dataset \cite{Arora_2019, Gayathri_2020, Xu_2017, Li_2017}. Modified versions of the original images can be obtained with rotation, flipping, or random cropping techniques \cite{Shorten_2019}. The limitation of these techniques is that the diversity of the resulting dataset is limited \cite{Lim_2020, Zhou_2020}. An alternate technique is the generation of synthetic imagery using GANs to augment the training data.

\subsection{GAN-based Approaches to Addressing Data Imbalance for DR}

Medical datasets are often imbalanced \cite{Rahman_2013, DerChiang_2010}, as such, there has been extensive work carried out on the generation of synthetic medical imagery (see Table \ref{tab:GansMedicalImages}). The quality and diversity of synthetic medical imagery are evaluated manually by physicians and quantitatively with evaluation metrics such as SSIM, FID, etc. It can be seen in Table \ref{tab:GansMedicalImages} that there is no consensus for evaluating the quality and diversity of generated imagery. 

However, the quality of synthetic images is evaluated more significantly than diversity measures. The diversity evaluation is important as it indicates the degree of mode collapse, a potential problem of training GANs consisting in generating similar synthetic images for diverse input images.

In this work, these perspectives have acted as motivating factors, for enabling a more transparent assessment of a GAN's capacity to generate suitably diversified retinal fundus images with PDR.

To demonstrate the benefits of generating synthetic imagery, in several works the training dataset is augmented with the synthetic images, and classification performance is calculated with evaluation metrics like accuracy, sensitivity, precision, kappa, and specificity.

In \cite{Zhou_2020} a conditional GAN (CGAN) is used to generate retinal fundus images for each DR stage. Quality is evaluated using three methods: manual evaluation, FID, and Sliced Wasserstein distance (SWD). Five hundred real and five hundred synthetic images are mixed, and two experiments are undertaken. Three ophthalmologists labeled each image as real or synthetic and assigned a severity level of DR. FID and SWD were calculated between the real and synthetic images.

In \cite{Balasubramanian_2020} a DCGAN is used to generate retinal fundus images of PDR. Quality and diversity are evaluated using an average CD. The synthetic images are added to the training dataset. The InceptionV3 model \cite{Szegedy_2016} pre-trained on the ImageNet database \cite{Deng_2009} extracts features from the images with PDR from the augmented dataset. The CDs between the extracted features are calculated and their average is compared to the average of the CDs between the features extracted only from the real images with PDR.

In \cite{Lim_2020} a MixGAN is proposed based on progressive layers and Style transfer to generate retinal fundus images of different DR stages such as moderate NPDR, severe NPDR, and PDR. It is not specified how synthetically generated images were evaluated.

\section{Methodology}

In this work, the suitability of evaluation metrics for assessing the quality and diversity of GAN-based retinal fundus synthetic images is analyzed. For this purpose, the methodology of GANs architectures, dataset, evaluation metrics, and the proposed idea of identifying suitable evaluation metrics is discussed as follows:

\subsection{DCGAN Architecture}

It is important to understand the characteristics of GANs so that these models can easily be reimplemented and fine-tuned for generating high-quality synthetic images \cite{wang2021generative}. For generating synthetic images, DCGAN \cite{Radford_2016} is considered a baseline model due to its simple architecture. DCGAN architecture can easily be adopted and reimplemented for any type of imagery to address the data imbalance problem \cite{huang2021enhanced}. Therefore, this work adopted DCGAN for synthesizing PDR fundus images.

\begin{figure}[htp!]
    \centering
    \includegraphics[width=1\textwidth]{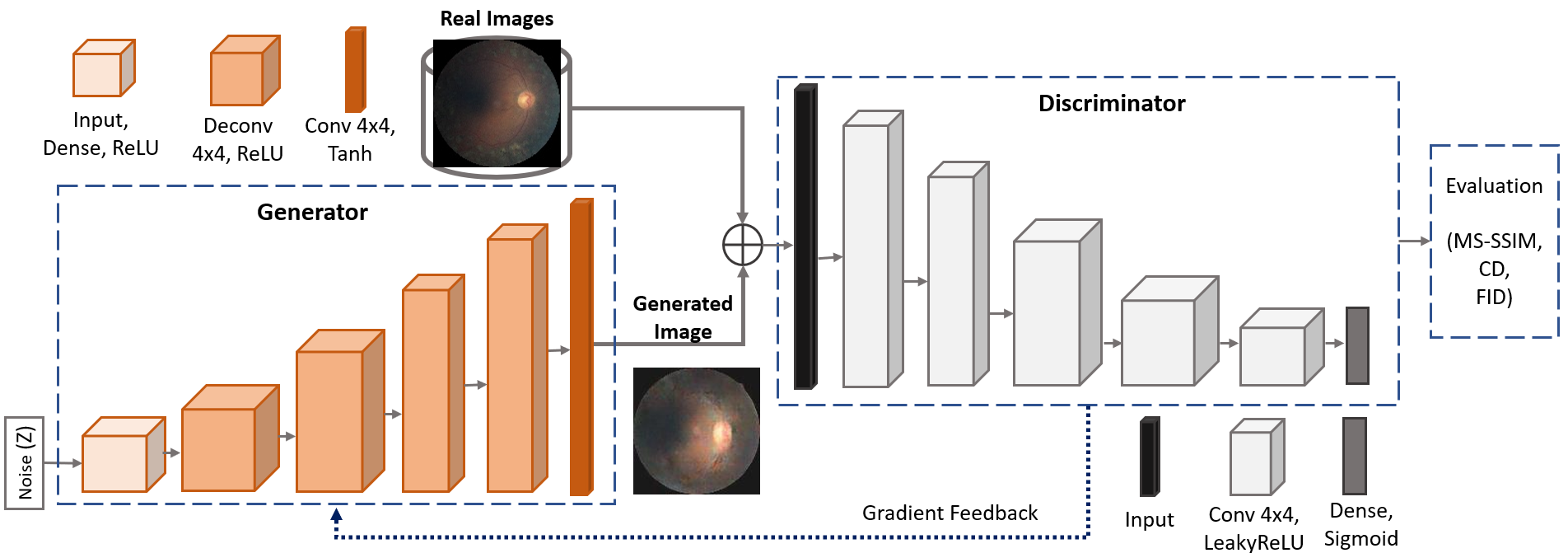}
    \caption{Architecture of DCGAN for PDR image synthesis. DCGAN generates synthetic PDR images using five convolutional layers in the discriminator and four deconvolutional layers in the generator.}
    \label{Fig.dcgan_final}
\end{figure}

The architecture of DCGAN is depicted in Fig. \ref{Fig.dcgan_final}. Initially, the DCGAN from \cite{Balasubramanian_2020} was adopted and reimplemented using the same parameter settings for generating PDR images. The DCGAN produced noisy images and was unable to generate realistic synthetic images. This could be due to batch normalization layers with upsampling and convolution layers in the generator and discriminator models of the DCGAN. Sometimes, batch normalization layers cannot reduce overfitting but explode gradients due to the usage of rescaling layers repeatedly \cite{kurach2019large}. So, the layers of the generator and discriminator models were redesigned with deconvolution and convolution layers only as reported in \cite{saad2022addressing}.

The DCGAN is fine-tuned with a learning rate of 0.0001 for the Adam optimizer. A Gaussian latent random value of 100 is used for input $z$ to the generator as adopted in \cite{Balasubramanian_2020} \cite{Chen_2019} \cite{Islam_2020}. The DCGAN is trained with a batch size of 16 for 500 epochs because both generator and discriminator models converge to a balanced state at this stage.

\subsubsection{Generation of Synthetic Images}
In GANs, synthetic images are generated without directly observing real data \cite{salimans2018improving}. Real images are used for training the discriminator, while the generator generates the synthetic images. During the training of GANs \cite{wang2021generative}, the generator model takes input from random values and generates noisy synthetic images. These images are passed to the discriminator. The discriminator also takes real images as input and distinguishes them from synthetic images. The discriminator backpropagates its feedback as gradients to the generator model. The generator model learns from that feedback and ideally enhances its capacity to generate realistically looking synthetic images. Once the generator is well-trained, it can generate several synthetic images by taking random input values. This is a baseline methodology used in GAN architectures to generate synthetic images.

\subsection{Retinal Fundus Imagery}
Publicly available retinal fundus imagery is imbalanced and increasingly limited as the severity of DR increases. As denoted in Table \ref{tab:datasets}, PDR is the minority class associated with DR. This work focuses on generating imagery of a minority class representative of the PDR.

In this work, the Kaggle dataset \cite{Cuadros_2009} is used, which is the largest publicly available dataset containing retinal fundus images from patients with different DR stages. Table \ref{tab:Distribution of images for every class, from the Kaggle training and test datasets} denotes the number of images per stage of DR available in the training and test datasets. The distribution of retinal fundus images within the classes varies significantly across training and test datasets.

\begin{table}[hbt!]
\centering
\caption{\textbf{Image distributions of different DR stages in the Kaggle dataset~\cite{Cuadros_2009}.}}
\begin{tabular}{p{1cm}p{2.3cm}p{1cm}p{1.6cm}p{2.3cm}p{1.8cm}p{0.6cm}}
\toprule
 & \textbf{Total No. Images} & \textbf{No DR} & \textbf{Mild NPDR} & \textbf{Moderate NPDR} & \textbf{Severe NPDR} & \textbf{PDR} \\
\midrule
Train & 35126 & 25810 & 2443 & 5292 & 873 & 708 \\
Test & 53576 & 39533 & 3762 & 7861 & 1214 & 1206 \\
\bottomrule
\end{tabular}
\label{tab:Distribution of images for every class, from the Kaggle training and test datasets}
\end{table}

PDR images were rescaled to a resolution of 128 x 128 for training the DCGAN. Generally, the DCGAN model works with a resolution of 128 x 128 as it is considered an intermediate size which is not too low to degrade the pixel information and not too high such that it is difficult to handle the training of the DCGAN \cite{Diaz-Pinto_2019}. In the domain of biomedical imagery, this resolution is commonly adopted to train GANs for generating synthetic images \cite{Zhou_2020} \cite{Balasubramanian_2020} \cite{Islam_2020} \cite{Diaz-Pinto_2019} \cite{saad2022addressing}.

For the classification of PDR images, the size of the images depends upon the type of classifier. A complex classifier, such as variants of CNNs, requires high-resolution images for learning salient image features. In contrast, a traditional classifier, such as a support vector machine (SVM) \cite{Cristianini_2000}, can efficiently work with lower image resolutions. In this work, images were rescaled to a resolution of 227 x 227 for the CNN model \cite{Gayathri_2020} and 224 x 224 for the EfficientNet model \cite{tan2019efficientnet}.

\subsubsection{Selection of Images for the Classifiers and GANs}

When training a multi-class classifier, images from each class are required to be diversified significantly as compared to the images of alternate classes. Therefore, it is essential to analyze the distribution of all images to decide the selection of diversified images for the classifier. To that end, MS-SSIM, CD, and FID are used to evaluate the similarity of images of each class to alternate classes for the training and test datasets as indicated in Table \ref{tab:evalMetricsTrainingAndTestDatasets}.  

To find the correlation among images of different classes from the training dataset and test dataset, 708 image samples from the training dataset and 1206 image samples from the test dataset were randomly selected for each class to measure the similarity scores. The 708 and 1206 samples were selected because these are the upper bounds that can be used for all classes. Table \ref{tab:evalMetricsTrainingAndTestDatasets} indicates that the distribution of images of all classes follows a relatively similar pattern in the training as compared to the test sets. It shows that the existing distribution of images is significant and should be used.

This work focuses on augmenting the PDR class using GAN-based synthetic images. Therefore, 708 images of PDR are used for training the DCGAN.

\subsection{Evaluation of GAN-based Synthetic Imagery}
\label{sec:evalMetrics}

\subsubsection{MS-SSIM}
The MS-SSIM metric evaluates the quality and diversity of synthetic images using perceptual similarities between images. It computes the similarity between images based on pixels and structural information of images. A higher value of MS-SSIM indicates higher similarity while a lower value of MS-SSIM indicates higher diversity between images of a single class \cite{odena2017conditional}. MS-SSIM is measured between two images \textit{a} and \textit{b} using Eq. \ref{eq:ms-ssim}.
\begin{equation}
\operatorname{MS}-\operatorname{SSIM}(a, b)=I_M(a, b)^{\alpha_M} \prod_{j=1}^M C_j(a, b)^{\beta_j} S_j(a, b)^{\gamma_j}\label{eq:ms-ssim}
\end{equation}
In Eq. \ref{eq:ms-ssim} \cite{Borji_2019}, structure (\textit{S}) and contrast (\textit{C}) image features are computed using the \textit{j} scale. \textit{M} indicates the coarsest scale for measuring luminance (\textit{I}). Weight parameters such as $\alpha$ $\beta$, and $\gamma$ are used for measuring \textit{S}, \textit{C}, and \textit{I} values.     

In this work, 708 image pairs (real-synthetic) are selected randomly from real and synthetic datasets to measure the MS-SSIM score for the quality of synthetically generated images. However, 354 image pairs (real-real) from the real dataset and 354 image pairs (synthetic-synthetic) from the synthetic dataset are selected randomly to measure the MS-SSIM scores for the diversity of synthetically generated images.
\subsubsection{CD}
CD is used to assess the quality and diversity of images. The cosine distance is computed by extracting the feature vectors of images using deep neural networks \cite{salimans2018improving}. CD for two images is computed using the feature vectors \textit{f1} and \textit{f2} as defined in Eq. \ref{fig:CosDistFormula}.  
\begin{equation}
\centering
\label{fig:CosDistFormula}
CD (\textit{f1}, \textit{f2}) = 1 - \frac{\vec{\textit{f1}} \cdot \vec{\textit{f2}}}{||\vec{\textit{f1}}|| \times ||\vec{\textit{f2}}||}
\end{equation}
In Eq. \ref{fig:CosDistFormula} \cite{salimans2018improving}, \textit{f1} and \textit{f2} refer to the feature vectors extracted from 2 images. A higher value of CD indicates higher diversity between images of a single class. In this work, feature vectors are extracted from an InceptionV3 model pre-trained on the ImageNet dataset. The CD is computed using 708 real and 708 synthetic images.

\subsubsection{FID}
FID is an evaluation metric used for assessing the quality and diversity of synthetic images. It computes the quality of images using the Wasserstein-2 distance between real and synthetic images. FID uses an Inception-V3 model pre-trained on the ImageNet dataset to measure the distance \cite{Borji_2019}. FID is computed between two sets of images \textit{x} and \textit{y} as defined in Eq. \ref{fig:fidFormula}.  
\begin{equation}
\label{fig:fidFormula}
FID (x, y) = ||m_1 - m_2||^2 + Tr(C_1 + C_2 – 2 \cdot \sqrt{C_1 \times C_2})
\end{equation}
In Eq. \ref{fig:fidFormula} \cite{Borji_2019}, \textit{$m_1$} and \textit{$m_2$} denote the vectors containing the mean of every feature from sets of images \textit{x} and \textit{y}, respectively. However, \textit{Tr} indicates a trace representing the sum of the elements from the main diagonal of a matrix. \textit{$C_1$ and $C_2$} represent covariance matrices for the feature vectors from the sets of images \textit{x} and \textit{y}, respectively.

In this work, 708 real and 708 synthetic images are selected to measure the FID score. A lower value of FID indicates a higher quality of synthetic images as compared to real images. 

\subsection{Normalization of Evaluation Metrics}

The evaluation metrics MS-SSIM, CD, and FID are computed differently using perceptual features and distance-based measures. These evaluation metrics have different scales for similarity measures. The resultant values are presented in a non-uniform manner such that higher MS-SSIM scores indicate higher similarity while higher CD and FID distance scores indicate lower similarity. 

Therefore, it is significant to normalize these metrics so that all these metrics evaluate the quality and diversity of synthetic images using similarity measures with a uniform scale. For this purpose, Eq. \ref{fig:normalization MS-SSIM}, \ref{fig:normalization cos dist}, and \ref{fig:normalization FID} are proposed.

\begin{equation}
\label{fig:normalization MS-SSIM}
    Normalized\; MS-SSIM = \frac{(MS-SSIM) - min\; (MS-SSIM)}{max\; (MS-SSIM) - min~(MS-SSIM)}
\end{equation}

\begin{equation}
\label{fig:normalization cos dist}
    Normalized\;CD = 1\;-\frac{CD\; - \;min\;CD} {max\;CD - min\;CD}
\end{equation}

\begin{equation}
\label{fig:normalization FID}
    Normalized\;FID = 1\;-\frac{FID\;-\;min\; FID}{max\;FID - min\;FID}
\end{equation}

The returned values are normalized to a 0 to 1 range. A high similarity between two sets of images is indicated by high values of the normalized evaluation metrics.

In Eq. \ref{fig:normalization MS-SSIM}, \ref{fig:normalization cos dist}, and \ref{fig:normalization FID}, \textit{max} MS-SSIM, \textit{max} CD and \textit{max} FID indicate the highest MS-SSIM, CD, and FID values respectively between two sets of images from the dataset. Similarly, \textit{min} MS-SSIM, \textit{min} CD, and \textit{min} FID indicate the lowest MS-SSIM, CD, and FID values respectively between two sets of images from the dataset.

Normalization is performed individually for each metric, each dataset (training and test), and each experiment. The normalized results obtained in different experiments or for different evaluation metrics are therefore not comparable.

\subsection{Classification of PDR Images}

\begin{figure}[htp!]
    \centering
    \includegraphics[width=1\textwidth]{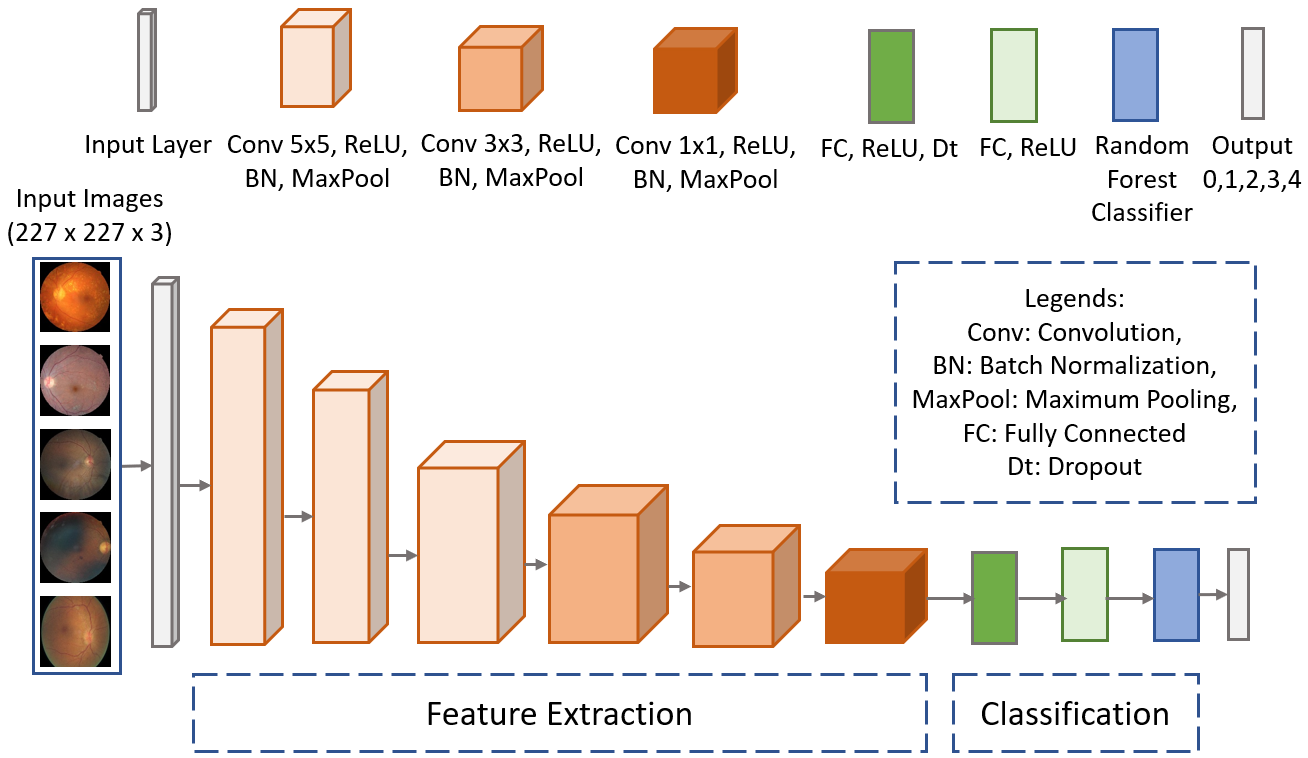}
    \caption{Classification of PDR images using CNN model.}
    \label{Fig.classifier}
\end{figure}
To assess the capacity of synthetically generated PDR images by DCGAN, these images are used to augment the minority class of PDR in the imbalanced dataset of \cite{Cuadros_2009}. Augmentation is undertaken in order to improve the classification performance of PDR disease. To this end, a state-of-the-art multi-class classifier from \cite{Gayathri_2020} is reimplemented for comparing the PDR classification results. The architecture of the classifier is designed with a CNN for feature extraction and a Random Forest model for classification as depicted in Fig. \ref{Fig.classifier}. In \cite{Gayathri_2020}, a single training dataset \cite{Cuadros_2009} is used. The CNN classifier is trained with 10-fold cross-validation using a batch size of 64 and stochastic gradient descent (SGD) optimizer with a learning rate of 0.003 as reported in \cite{Gayathri_2020}. In this work, a CNN is trained with 10-fold cross-validation using the same approach as in \cite{Gayathri_2020} for all the training dataset classes except PDR. For the PDR class, PDR images from the test dataset \cite{Cuadros_2009} are used for 10-fold cross-validation. This approach is used to avoid any biasing as the PDR images from the training dataset were used for training the DCGAN for generating synthetic images. The CNN classifier required a high computational time of approximately hundreds of hours for training the whole dataset. Therefore, CNN is trained on one batch of images from each fold only. However, an additional classifier such as EfficientNet \cite{tan2019efficientnet} is also trained with 20 epochs using pretrained weights of the ImageNet dataset \cite{Deng_2009} on the whole dataset including all batches of images. The EfficientNet classifier required a low computational cost of 9 minutes to train all batches of images in the dataset.  

In \cite{Gayathri_2020}, the training dataset is highly imbalanced and there is no discussion of alleviating bias in the trained model. Consequently, this work uses class weights in the loss computation to address potential bias in the model. Class weights penalize the classifier for the instances that are misclassified. Class weight values are selected using a formula defined in Eq. \ref{eq:class weights} \cite{tensorflow_classweights}, corresponding to the ``balanced'' value of the \emph{class\_weight} parameter from the scikit-learn library \cite{scikitlearn_classweights}.

Classification performance is evaluated with the $F_1$ score and Area Under the Curve (AUC), using the formulas denoted in Eq. \ref{eq:F1 score} and Eq. \ref{eq:AUC OVR}. $F_1$ score is calculated based on recall and precision. The recall of class PDR indicates the proportion of patients with PDR that were diagnosed as having PDR. The precision of class PDR indicates the proportion of patients diagnosed with PDR that have PDR. It is important that patients with PDR are diagnosed correctly, in order to get the treatment required. It is also important that patients are only diagnosed with PDR if they have the disease, in order to prevent unnecessary/incorrect treatments. AUC measures how accurately the border between the classes is identified by the classifier. The one-versus-rest (OVR) approach \cite{Provost_2000} is used for calculating AUC.

\begin{equation}
\label{eq:class weights}
    Class\;weight\;for\;class\;x = \frac{Total\;no.\;of\;images\;of\;all\;classes}{Total\;no.\;of\;classes \times No.\;of\;images\;of\;class\;x}
\end{equation}

\begin{equation} 
    \label{eq:F1 score}
    \text{$F_1$} = 2 \times \frac{precision \times recall}{precision + recall}
\end{equation}

\begin{equation} 
    \label{eq:recall}
    \text{$recall$} = \frac{TP}{TP + FN}
\end{equation}

\begin{equation} 
    \label{eq:precision}
    \text{precision} = \frac{TP}{TP + FP}
\end{equation}

\begin{equation} 
    \label{eq:AUC OVR}
    \text{AUC}_{\text{OVR}} = \frac{1}{c} \sum _{i=0}^{c-1}{\text{AUC}}(C_i, C_i^C)
\end{equation}

The $F_1$ score, recall and precision are calculated separately for each class. In Eq. \ref{eq:recall} and Eq. \ref{eq:precision}, \textit{TP} denotes the true positives. \textit{TP} of class \textit{x} represents the number of images belonging to class \textit{x} that are classified correctly. \textit{FN} denotes the false negatives. \textit{FN} of class \textit{x} represents the number of images belonging to class \textit{x} that are classified as belonging to other classes. \textit{FP} denotes the false positives. \textit{FP} of class \textit{x} represents the number of images classified as class \textit{x}, that belong to other classes.

\subsection{Correlation of Quality, Diversity, and Classification Performances}

The synthetically generated PDR images are used to augment the dataset for improving the classification performance of the classifier. The classification performance is higher when the training images are of high quality. Therefore, a metric is considered suitable for evaluating the quality of synthetic imagery if a high-quality score indicated by that metric correlates with a high classification performance. If the classification performance is lower with high-quality scores, then it shows that either the quality evaluation metric does not correctly assess the quality of images or the images still lack the level of quality to improve the classifier's score. Similarly, classification performance is higher when the training images are diverse. Therefore, a metric is considered suitable for evaluating the diversity of synthetic imagery if a high diversity score indicated by that metric aligns with a high classification performance. A lower classification score with highly diversified images indicates that either the evaluation metric does not correctly assess the diversity of images or the desired level of diversity is not achieved. Therefore, the quality and diversity scores are analyzed using each evaluation metric and correlated against the classification performance. The intent is to assess the metrics' suitability for evaluating the quality and diversity of synthetic imagery.

\section{Results and Discussion}

\subsection{Critical Analysis of Quantitative Evaluation Metrics}

The similarity scores between retinal fundus images with different stages of DR are derived using MS-SSIM, CD, and FID, as denoted in Table \ref{tab:evalMetricsTrainingAndTestDatasets}. FID and MS-SSIM produce the most performant similarity scores, as values between images of the same class are greater than values between images of alternate classes. However, CD does not provide a utilizable similarity score as values derived for the images of one class are more similar to the images of alternate classes. 

\FloatBarrier
\begin{table}[hbt]
\centering
\caption{\textbf{Similarity between sampled retinal fundus images with different stages of DR measured via MS-SSIM, CD, and FID with 95\% confidence interval.}}
\begin{tabular}{p{1.1cm}p{1.6cm}p{1.6cm}p{1.6cm}p{1.6cm}p{1.6cm}p{1.6cm}}
\toprule
\multirow{2}{1cm}{\textbf{EM, Dataset}}
& \textbf{Class}
& \textbf{No DR}
& \textbf{Mild NPDR}
& \textbf{Mod. NPDR}
& \textbf{Sev. NPDR}
& \textbf{PDR}\\
\\
\midrule
\multirow{5}{1cm}{FID, \\ Training \\ data} & \multirow{5}{1.7cm}{No DR \\ Mild NPDR \\ Mod. NPDR \\ Sev. NPDR \\ PDR}

& \multirow{5}{0.6cm}{\textbf{1} \\ 0.389{\ssmall \;\textpm\;0.019} \\ 0.196{\ssmall\;\textpm\;0.09} \\ 0 \\ 0.054{\ssmall\;\textpm\;0.002}}

& \multirow{5}{0.65cm}{0.389{\ssmall\;\textpm\;0.019} \\ \boldmath \textbf{$>$0.999} \\ 0.359{\ssmall\;\textpm\;0.017} \\ 0.261{\ssmall\;\textpm\;0.013} \\ 0.110{\ssmall\;\textpm\;0.005}}

& \multirow{5}{0.65cm}{0.196{\ssmall\;\textpm\;0.009} \\ 0.359{\ssmall\;\textpm\;0.017} \\ \boldmath \textbf{$>$0.999} \\ 0.391{\ssmall\;\textpm\;0.019} \\ 0.226{\ssmall\;\textpm\;0.011}}

& \multirow{5}{0.65cm}{0 \\ 0.261{\ssmall\;\textpm\;0.013} \\ 0.391{\ssmall\;\textpm\;0.019} \\ \boldmath \textbf{$>$0.999} \\ 0.096{\ssmall\;\textpm\;0.004}}

& \multirow{5}{0.6cm}{0.054{\ssmall\;\textpm\;0.002} \\ 0.110{\ssmall\;\textpm\;0.005} \\ 0.226{\ssmall\;\textpm\;0.011} \\ 0.096{\ssmall\;\textpm\;0.004} \\ \boldmath \textbf{$>$0.999}}
\\
&&&&&& \\
&&&&&& \\
&&&&&& \\

\midrule
\multirow{5}{1cm}{FID, \\ Test \\ data} & \multirow{5}{1.7cm}{No DR \\ Mild NPDR \\ Mod. NPDR  \\ Sev. NPDR \\ PDR} 
& \multirow{5}{0.6cm}{\textbf{1} \\ 0.409{\ssmall\;\textpm\;0.020} \\ 0.253{\ssmall\;\textpm\;0.012} \\ 0.140{\ssmall\;\textpm\;0.007} \\ 0}

& \multirow{5}{0.65cm}{0.409{\ssmall\;\textpm\;0.020} \\ \boldmath \textbf{$>$0.999} \\ 0.390{\ssmall\;\textpm\;0.019} \\ 0.290{\ssmall\;\textpm\;0.014} \\ 0.023{\ssmall\;\textpm\;0.001}}

& \multirow{5}{0.65cm}{0.253{\ssmall\;\textpm\;0.012} \\ 0.390{\ssmall\;\textpm\;0.019} \\ \boldmath \textbf{$>$0.999} \\ 0.453{\ssmall\;\textpm\;0.022} \\ 0.142{\ssmall\;\textpm\;0.007}}

& \multirow{5}{0.65cm}{0.140{\ssmall\;\textpm\;0.007} \\ 0.290{\ssmall\;\textpm\;0.014} \\ 0.453{\ssmall\;\textpm\;0.022} \\ \boldmath \textbf{$>$0.999} \\ 0.102{\ssmall\;\textpm\;0.005}}

& \multirow{5}{0.6cm}{0 \\ 0.023{\ssmall\;\textpm\;0.001} \\ 0.142{\ssmall\;\textpm\;0.007} \\ 0.102{\ssmall\;\textpm\;0.005} \\ \boldmath \textbf{$>$0.999}}
\\
&&&&&& \\
&&&&&& \\
&&&&&& \\

\midrule
\multirow{5}{1cm}{CD, \\ Training \\ data} & \multirow{5}{1.7cm}{No DR \\ Mild NPDR \\ Mod. NPDR  \\ Sev. NPDR \\ PDR} 

& \multirow{5}{0.6cm}{0.543{\ssmall\;\textpm\;0.027} \\ 0.634{\ssmall\;\textpm\;0.032} \\ 0.341{\ssmall\;\textpm\;0.017} \\ \textbf{1} \\ 0.426{\ssmall\;\textpm\;0.021} }

& \multirow{5}{0.65cm}{0.634{\ssmall\;\textpm\;0.032} \\ 0.364{\ssmall\;\textpm\;0.018} \\ 0.118{\ssmall\;\textpm\;0.006} \\ 0.659{\ssmall\;\textpm\;0.033} \\ 0}

& \multirow{5}{0.65cm}{0.341\;\textpm{\ssmall\;0.017} \\ 0.118{\ssmall\;\textpm\;0.006} \\ 0.18{\ssmall\;\textpm\;0.009} \\ 0.714{\ssmall\;\textpm\;0.036} \\ 0.088{\ssmall\;\textpm\;0.004}}

& \multirow{5}{0.65cm}{ \textbf{1}\\ \textbf{0.659{\ssmall\;\textpm\;0.033}}\\
\textbf{0.714{\ssmall\;\textpm\;0.036}} \\ 0.818{\ssmall\;\textpm\;0.041} \\ \textbf{0.462{\ssmall\;\textpm\;0.023}}}

& \multirow{5}{0.6cm}{0.426{\ssmall\;\textpm\;0.021} \\ 0 \\ 0.088{\ssmall\;\textpm\;0.004} \\ 0.462{\ssmall\;\textpm\;0.023} \\ 0.318{\ssmall\;\textpm\;0.016}}
\\
&&&&&& \\
&&&&&& \\
&&&&&& \\

\midrule  
\multirow{5}{1cm}{CD, \\ Test \\ data} 
& \multirow{5}{1.7cm}{No DR \\ Mild NPDR \\ Mod. NPDR  \\ Sev. NPDR \\ PDR} 
& \multirow{5}{0.6cm}{\textbf{0.772{\ssmall\;\textpm\;0.039}} \\ \textbf{0.714{\ssmall\;\textpm\;0.036}} \\ \textbf{0.685{\ssmall\;\textpm\;0.034}} \\ 0.654{\ssmall\;\textpm\;0.033} \\ \textbf{0.58{\ssmall\;\textpm\;0.029}} }

& \multirow{5}{0.65cm}{0.714{\ssmall\;\textpm\;0.036} \\ 0.258{\ssmall\;\textpm\;0.013} \\ 0.387{\ssmall\;\textpm\;0.019} \\ 0.62{\ssmall\;\textpm\;0.031} \\ 0.013{\ssmall\;\textpm\;0.001} }

& \multirow{5}{0.65cm}{0.685{\ssmall\;\textpm\;0.034} \\ 0.387{\ssmall\;\textpm\;0.019} \\ 0.193{\ssmall\;\textpm\;0.01} \\ 0.624{\ssmall\;\textpm\;0.031} \\ 0 }

& \multirow{5}{0.65cm}{0.654{\ssmall\;\textpm\;0.033} \\ 0.62{\ssmall\;\textpm\;0.031} \\ 0.624{\ssmall\;\textpm\;0.031} \\ \textbf{1} \\ 0.474{\ssmall\;\textpm\;0.024} }

& \multirow{5}{0.6cm}{0.58{\ssmall\;\textpm\;0.029} \\ 0.013{\ssmall\;\textpm\;0.001} \\ 0 \\ 0.474{\ssmall\;\textpm\;0.024} \\ 0.45{\ssmall\;\textpm\;0.023} }
\\
&&&&&& \\
&&&&&& \\
&&&&&& \\

\midrule
\multirow{5}{1cm}{MS-SSIM, \\ Training \\ data} & \multirow{5}{1.7cm}{No DR \\ Mild NPDR \\ Mod. NPDR  \\ Sev. NPDR \\ PDR} 

& \multirow{5}{0.6cm}{\textbf{1} \\ 0.02{\ssmall\;\textpm\;0.001} \\ 0.04{\ssmall\;\textpm\;0.002} \\ 0.039{\ssmall\;\textpm\;0.002} \\ 0.018{\ssmall\;\textpm\;0.001}}

& \multirow{5}{0.65cm}{0.02{\ssmall\;\textpm\;0.001} \\ \textbf{1} \\ 0.036{\ssmall\;\textpm\;0.002} \\ 0.053{\ssmall\;\textpm\;0.003} \\ 0 \\ }

& \multirow{5}{0.65cm}{0.04{\ssmall\;\textpm\;0.002}\\ 0.036{\ssmall\;\textpm\;0.002} \\ \textbf{1} \\ 0.029{\ssmall\;\textpm\;0.001} \\ 0.009\\}

& \multirow{5}{0.65cm}{0.039{\ssmall\;\textpm\;0.002} \\ 0.053{\ssmall\;\textpm\;0.003} \\ 0.029{\ssmall\;\textpm\;0.001} \\ \textbf{1} \\ 0.003\\}

& \multirow{5}{0.6cm}{0.018{\ssmall\;\textpm\;0.001} \\ 0 \\ 0.009 \\ 0.003 \\ \textbf{1}\\}
\\
&&&&&& \\
&&&&&& \\
&&&&&& \\

\midrule
\multirow{5}{1cm}{MS-SSIM, \\ Test \\ data} 
& \multirow{5}{1.7cm}{No DR \\ Mild NPDR \\ Mod. NPDR  \\ Sev. NPDR \\ PDR} 
 
& \multirow{5}{0.6cm}{\textbf{1\\} 0.009\\ 0.01{\ssmall\;\textpm\;0.001} \\ 0.021{\ssmall\;\textpm\;0.001} \\ 0.002 \\}

& \multirow{5}{0.65cm}{0.009 \\ \textbf{1} \\ 0 \\ 0.007 \\ 0.003\\}

& \multirow{5}{0.65cm}{0.01{\ssmall\;\textpm\;0.001} \\ 0 \\ \textbf{1} \\ 0.015{\ssmall\;\textpm\;0.001} \\ 0.014{\ssmall\;\textpm\;0.001} }

& \multirow{5}{0.65cm}{0.021{\ssmall\;\textpm\;0.001} \\ 0.007 \\ 0.015{\ssmall\;\textpm\;0.001} \\ \textbf{1} \\ 0.009 \\ }

& \multirow{5}{0.6cm}{0.002 \\ 0.003 \\ 0.014{\ssmall\;\textpm\;0.001} \\ 0.009 \\ \textbf{1} \\ }
\\
&&&&&& \\
&&&&&& \\
&&&&&& \\

\bottomrule
\multicolumn{7}{l}{EM: Evaluation Metric; Mod: Moderate; Sev: Severe}
\end{tabular}
\label{tab:evalMetricsTrainingAndTestDatasets}
\end{table}
\FloatBarrier

Consequently, FID and MS-SSIM are considered suitable metrics for evaluating synthetic images, if these images are representative of the real images of the targeted class. The quantitative measures of FID and MS-SSIM enable the selection of suitable synthetic imagery for data augmentation when training a classifier.

CD is unsuitable for evaluating synthetic imagery representative of its class because it is calculated based on the angle between the extracted features. If the angle between two feature vectors is zero, they are not necessarily identical.

\subsection{Evaluation of Synthetic PDR Imagery} 

\subsubsection{Quality of Synthetic PDR Imagery}

\begin{figure}[htp!]
    \centering
    \includegraphics[width=1\textwidth]{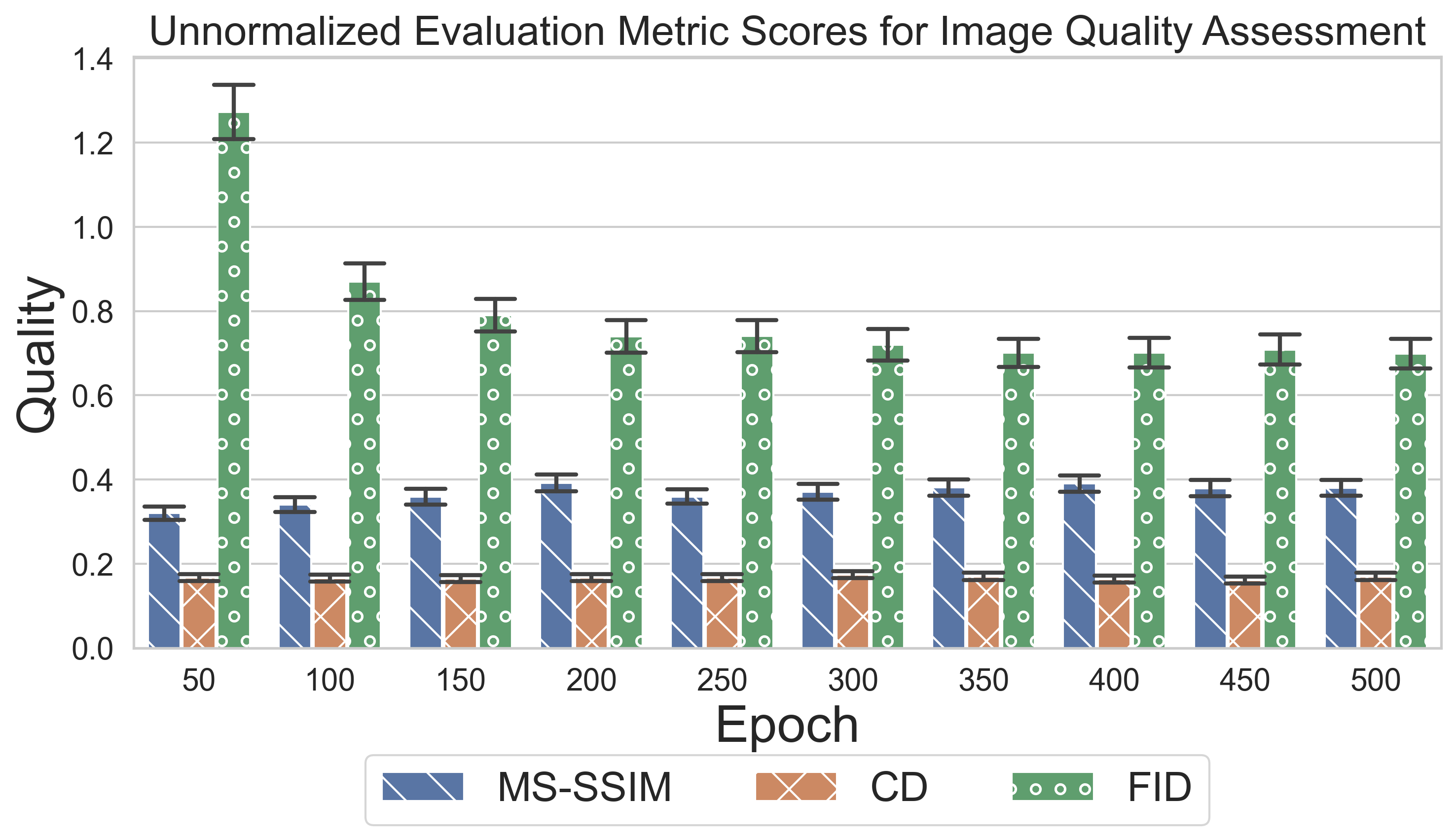}
    \caption{Quality scores for each evaluation metric indicate the comparison of quality between synthetic and real imagery. Quality is evaluated using unnormalized scores of MS-SSIM, CD, and FID metrics.}
    \label{Fig.qualitywithoutnorm}
\end{figure}

\begin{figure}[htp!]
    \centering
    \includegraphics[width=1\textwidth]{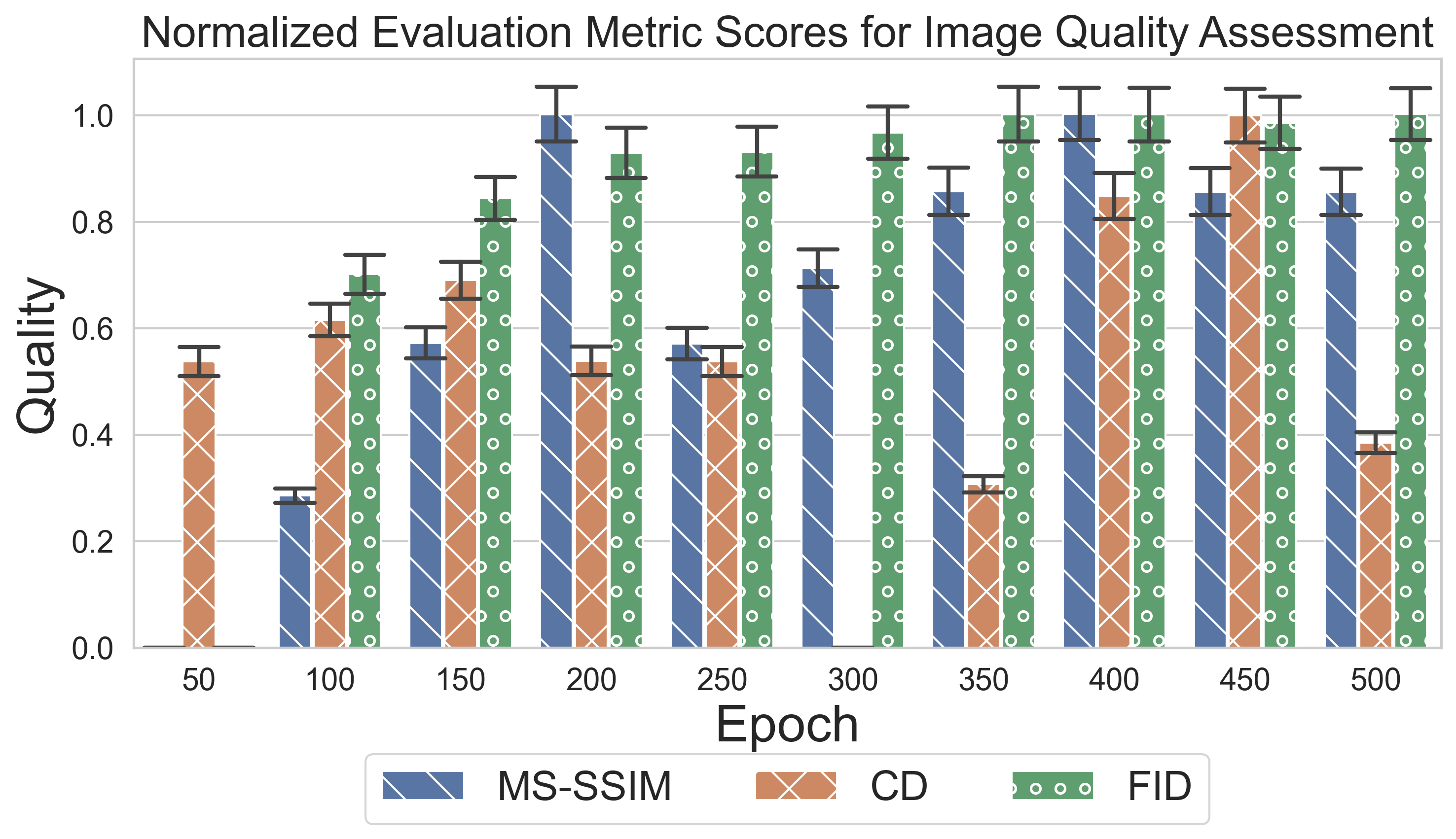}
    \caption{Quality scores for each evaluation metric indicate the comparison of quality between synthetic and real imagery. Quality is evaluated using normalized scores of MS-SSIM, CD, and FID metrics.}
    \label{Fig.qualitywithnorm}
\end{figure}

The quality of synthetic images generated is calculated via the evaluation metrics MS-SSIM, CD, and FID (see Section \ref{sec:evalMetrics}). The quality of synthetic images is evaluated using the unnormalized and normalized values of MS-SSIM, CD, and FID as depicted in Fig. \ref{Fig.qualitywithoutnorm} and Fig. \ref{Fig.qualitywithnorm}. An improvement in quality with unnormalized metrics values is denoted as follows: higher MS-SSIM, lower CD, and lower FID when compared synthetic to real images as depicted in Fig. \ref{Fig.qualitywithoutnorm}. In Fig. \ref{Fig.qualitywithoutnorm}, the decrease in FID indicates that the imagery generated in the last epochs of training is of higher quality than that generated in the early epochs. MS-SSIM indicates an improvement in quality up to epoch 200 and then oscillates with inconsistent behavior. CD indicates consistent behavior in the quality of images generated throughout training.

An improvement in quality with normalized metrics values is denoted as follows: higher MS-SSIM, higher CD, and higher FID when comparing synthetic to real images as depicted in Fig. \ref{Fig.qualitywithnorm}. In Fig. \ref{Fig.qualitywithnorm}, normalized FID indicates an increase in the quality indicating that the last epochs of training have higher-quality images. The normalized MS-SSIM and normalized CD indicate a significant improvement in quality for the first few epochs and then oscillate inconsistently.

FID is used to evaluate the quality of synthetic images as compared to real images significantly as evidenced by FID analysis in Fig. \ref{Fig.qualitywithoutnorm} and Fig. \ref{Fig.qualitywithnorm}. MS-SSIM and CD do not enable meaningful analysis of the quality of images as they measure the similarity and distance between image pairs respectively. Synthetic images may have different statistical properties than real images, such as different color distributions or noise characteristics, which can lead to a lower MS-SSIM score even if the synthetic images are of high quality. Similarly, CD does not consider the spatial relationships between the pixels in the image, which can be important for the overall visual quality between real and synthetic images. Therefore, MS-SSIM and CD are not suitable metrics for evaluating the quality of synthetic images.

\subsubsection{Diversity of Synthetic PDR Imagery}

\begin{figure}[htp!]
    \centering
    \includegraphics[width=1\textwidth]{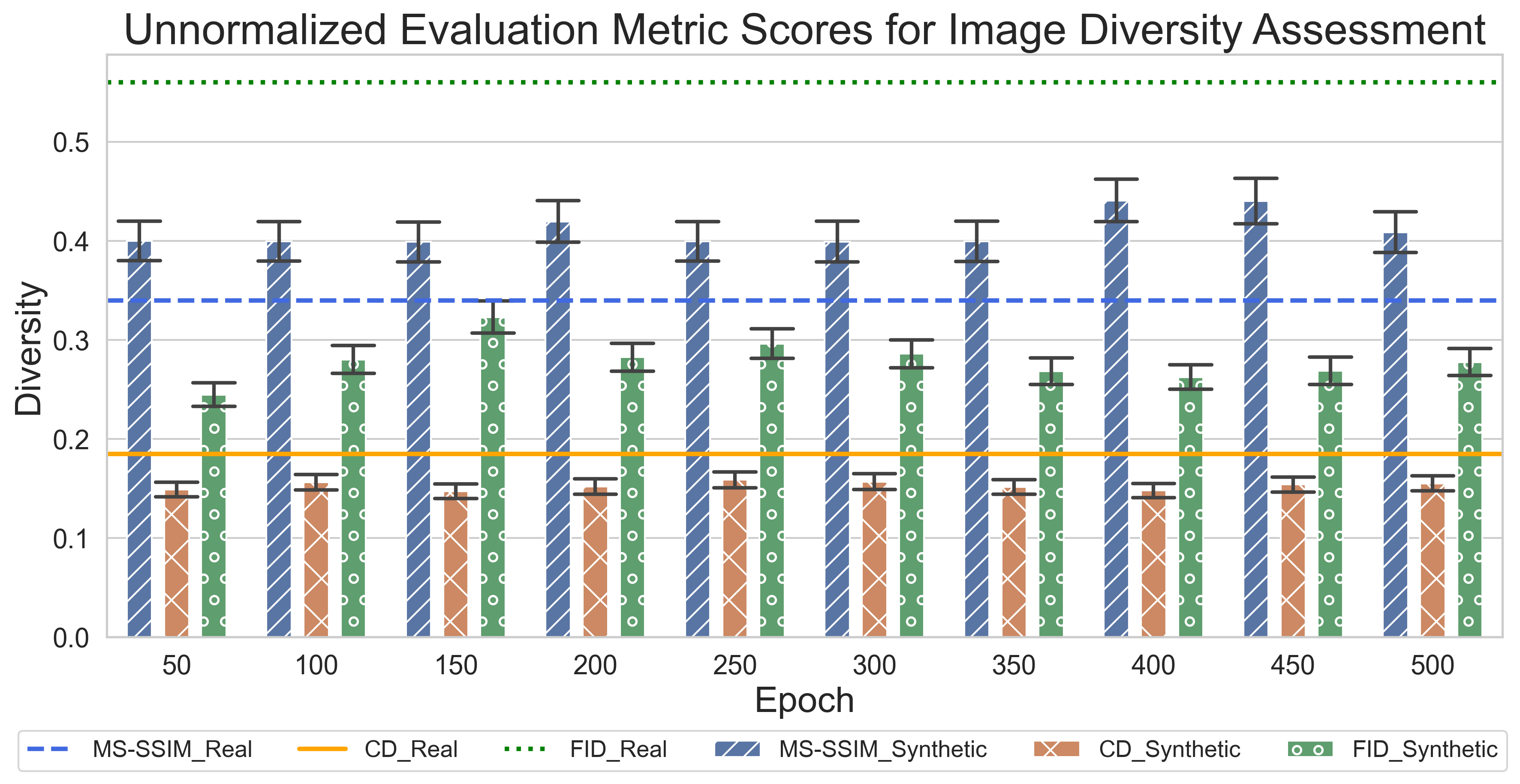}
    \caption{Diversity scores for each evaluation metric indicate the comparison of diversity between (synthetic:synthetic) and (real:real). Diversity is evaluated using unnormalized scores of MS-SSIM, CD, and FID metrics.}
    \label{Fig.diversitywithoutnorm}
\end{figure}

\begin{figure}[htp!]
    \centering
    \includegraphics[width=1\textwidth]{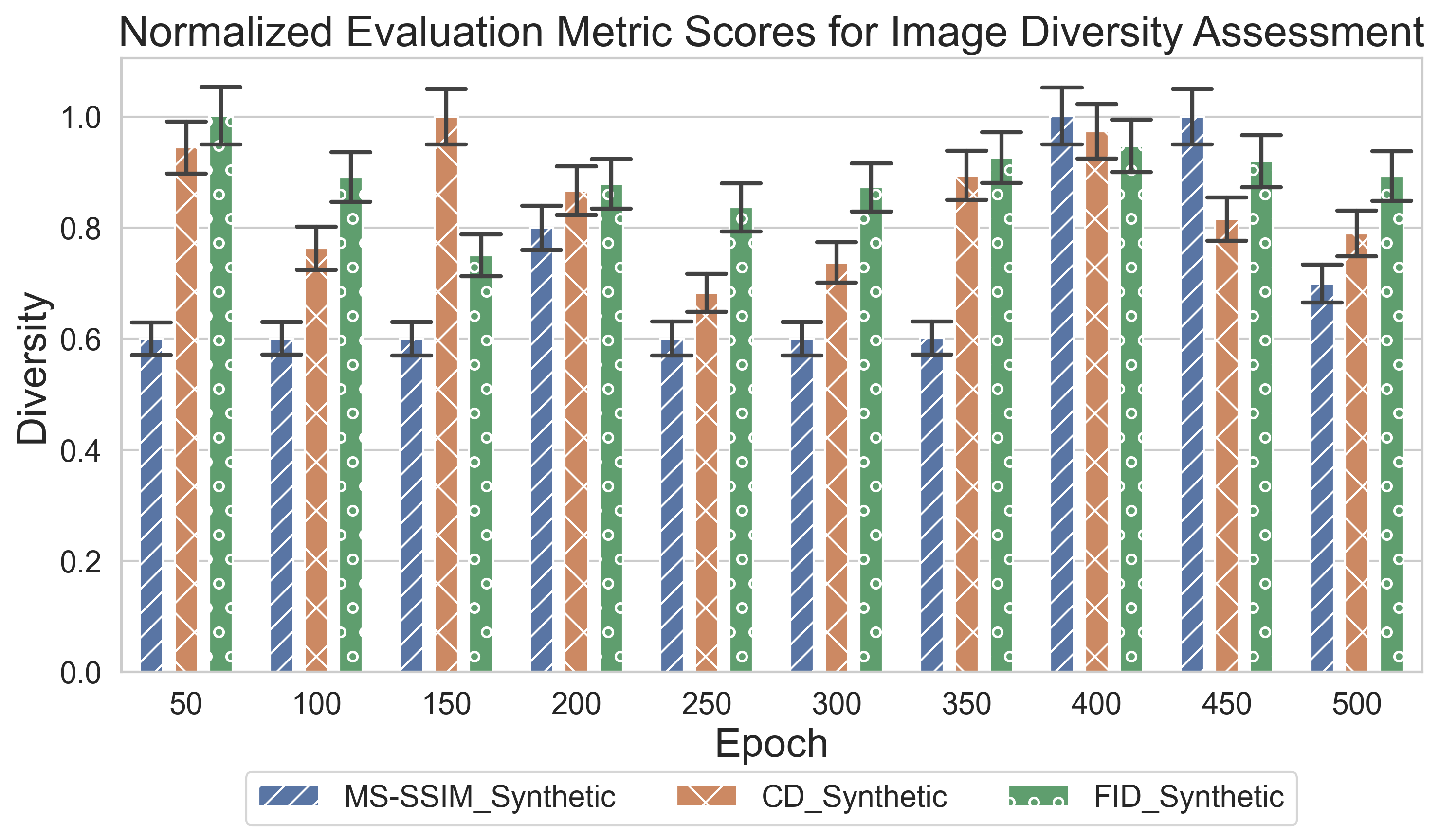}
    \caption{Diversity scores for each evaluation metric indicate the comparison of diversity between (synthetic:synthetic) and (real:real). Real images have higher diversity (metric scores=0). Diversity is evaluated using normalized scores of MS-SSIM, CD, and FID metrics.}
    \label{Fig.diversitywithnorm}
\end{figure}

The diversity of synthetic images generated is calculated via the evaluation metrics MS-SSIM, CD, and FID (see Section \ref{sec:evalMetrics}). The diversity of synthetic images is evaluated using the unnormalized and normalized values of MS-SSIM, CD, and FID as depicted in Fig. \ref{Fig.diversitywithoutnorm} and Fig. \ref{Fig.diversitywithnorm}. An improvement in diversity unnormalized metrics values is denoted as follows: lower MS-SSIM, higher CD, and higher FID when comparing synthetic to real images as depicted in Fig. \ref{Fig.diversitywithoutnorm}.

In Fig. \ref{Fig.diversitywithoutnorm}, FID indicates a significant drop in diversity throughout the training of the DCGAN. MS-SSIM indicates relatively consistent diversity for synthetic images until epoch 350. The diversity of synthetic images starts decreasing from epoch 350 to 400 and then improving from epochs 400 and onwards. The CD indicates consistent behavior of diversity of synthetic images compared to real images throughout the training of the DCGAN.

An improvement in diversity with normalized metric values is denoted as follows; lower MS-SSIM, lower CD, and lower FID when comparing synthetic to real images as depicted in Fig. \ref{Fig.diversitywithnorm}. In Fig. \ref{Fig.diversitywithnorm}, all three metrics indicate inconsistent behavior for the diversity of synthetic images as compared to real images. MS-SSIM indicates that real images are more diverse than synthetic images at various epochs because synthetic images lack the distribution of structure features as compared to real images. The CD indicates that real imagery is more diverse than most sets of synthetic imagery because the features extracted from real imagery are less dependent on each other. FID indicates that the features from the real imagery are spread over a larger area as it uses embedding layers of a pre-trained model.

\FloatBarrier
\FloatBarrier

MS-SSIM and CD metrics are used to evaluate the diversity of synthetic images as compared to real images. This work also depicts the significance of using these metrics to evaluate diversity, as depicted in Fig. \ref{Fig.diversitywithoutnorm} and Fig. \ref{Fig.diversitywithnorm}. FID is unsuitable for diversity evaluation as it provides inconsistent analysis for the diversity of synthetic images that is significantly away from the MS-SSIM and CD analysis.

\subsubsection{Selection of Synthetic Imagery for Augmenting Imbalanced Datasets}

The DCGAN-based synthetic images are ranked based on the quality and diversity scores measured by the MS-SSIM, CD, and FID evaluation metrics as indicated in Table \ref{tab:Ranking_table}. A significant variance is observed in the metric values for evaluating the quality and diversity of synthetic images. Therefore, it is important to find a suitable set of synthetic images that can be used for augmenting datasets. The ranking of synthetic images generated from each epoch enables the selection of synthetic images significantly, which helps in improving the performance of classifiers for augmented datasets. A rank of 1 indicates the most promising high-quality images while a rank of 7 indicates the lower-quality images. Similarly, a rank of 1 indicates higher diversity, while rank 10 indicates a lower diversity of synthetic images. 

\FloatBarrier

\begin{table}[htp!]
\centering
\caption{\textbf{Selection of DCGAN-based synthetic images based on quality and diversity ranking using MS-SSIM, CD, and FID metric scores to augment the original dataset. Bold values indicate the top-ranked and moderate-ranked scores.}
}

\begin{tabular}{p{2cm}p{2.2cm}p{0.6cm}p{0.6cm}p{0.6cm}p{0.6cm}p{0.6cm}p{0.6cm}p{0.6cm}p{0.6cm}p{0.6cm}p{0.6cm}}

\toprule

\textbf{Characteristic} & \textbf{Ranking Metric} & \multicolumn{10}{c}{\textbf{Epoch}} \\
& & 50 & 100 & 150 & 200 & 250 & 300 & 350 & 400 & 450 & 500 \\

\midrule

\multirow{2}{*}{Diversity} & MS-SSIM & 1 & 1 & 1 & \textbf{3} & \textbf{1} & 1 & 1 & 4 & 4 & \textbf{2} \\
& CD & 8 & 3 & 10 & \textbf{6} & \textbf{1} & 2 & 7 & 9 & 5 & \textbf{4} \\
&&&&&&&&&& \\
\multirow{1}{*}{Quality} & FID & 7 & 6 & 5 & \textbf{4} & \textbf{4} & 3 & 1 & 1 & 2 & \textbf{1} \\

\bottomrule

\end{tabular}
\label{tab:Ranking_table}
\end{table}
\FloatBarrier

The synthetic images with top-ranked and moderate-ranked scores are selected to assess the quality and diversity measures. FID scores at epochs 350, 400, and 500 achieved rank 1. However, synthetic images with epoch 500 are selected because this rank is also consistent with the best ranks of MS-SSIM and CD. Similarly, MS-SSIM and CD with epoch 250 achieved rank 1 scores. Therefore, synthetic images of epoch 250 are selected. The moderate-ranked quality and diversity scores of each epoch are also analyzed to select the synthetic images. The images of epoch 200 are selected as indicated by moderate-ranked scores in Table \ref{tab:Ranking_table}.

The detailed analysis of the synthetic images' best-ranked quality and diversity scores is compared using unnormalized and normalized metric scores as indicated in Table \ref{tab:Ranking_discussion_table}. In Table \ref{tab:Ranking_discussion_table}, a higher normalized FID score of 1 for epoch 500 indicates that the synthetic images preserve the best quality compared to the real images. In contrast, the moderate scores of MS-SSIM and CD for epoch 500 do not reflect the best diversity of synthetic images as compared to real images. Similarly, higher normalized values of MS-SSIM and CD for epoch 250 indicate that the synthetic images have the best diversity while the moderate score of FID at epoch 250 indicates the poor quality of synthetic images as compared to real images.  

PDR images contain several salient features such as the structure, shape, color, and size of blood vessels and lesions as depicted in Fig. \ref{Fig.real_synthetic}. It is important to learn and generate these features when synthesizing PDR images using GANs. In this work, DCGAN has generated synthetic PDR images that are representative of real images. FID has evaluated the quality of synthetic images compared to real images. In Table \ref{tab:Ranking_discussion_table}, the best quality of synthetic images is achieved at epoch 500 with an unnormalized value of 0.70 and the normalized value of 1. However, the suppressed structural features of vessels in synthetic PDR images are indicative of the DCGAN's limitation to generate high-quality synthetic images as depicted in Fig. \ref{Fig.real_synthetic}.

\FloatBarrier

\begin{table}[htp!]
\centering
\caption{\textbf{Comparing best-ranked DCGAN-based synthetic image datasets using unnormalized and normalized metric scores for quality and diversity measures.}
}

\begin{tabular}{p{0.8cm}p{0.6cm}p{1cm}p{0.7cm}p{0.6cm}p{1cm}p{0.7cm}p{0.6cm}p{1cm}p{0.7cm}p{2.4cm}}

\toprule

\textbf{Epoch} & \multicolumn{3}{c|}{\textbf{MS-SSIM}} & \multicolumn{3}{c|}{\textbf{CD}} & \multicolumn{3}{c|}{\textbf{FID}} & \textbf{Comment} \\
& Rank & Unnorm & Norm & Rank & Unnorm & Norm & Rank & Unnorm & Norm & \\

\midrule

200 & 3 & 0.42 & 0.8 & 6 & 0.152 & 0.868 & 4 & 0.74 & 0.93 & Moderate diversity, Moderate quality \\
250 & 1 & 0.4 & 0.6 & 1 & 0.159 & 0.684 & 4 & 0.74 & 0.93 & Higher diversity, Moderate quality \\
500 & 2 & 0.41 & 0.7 & 4 & 0.155 & 0.789 & 1 & 0.70 & 1 & Moderate diversity, Higher quality \\

\bottomrule
\multicolumn{11}{l}{MS-SSIM and CD values refer to the synthetic datasets.} \\
\end{tabular}
\label{tab:Ranking_discussion_table}
\end{table}
\FloatBarrier

\subsection{Assessment of Synthetic Imagery using Classification Scores}

Table \ref{tab:F1 score on training vs test dataset} indicates the classification scores of the CNN and EfficientNet classifiers using $F_1$ score and AUC score when trained on both the original and augmented datasets. Training the CNN was computationally expensive, taking several hours to train each iteration on the whole original dataset. The AUC scores of CNN and EfficientNet for the augmented dataset are improved compared to the original dataset as indicated in Table \ref{tab:F1 score on training vs test dataset}. $F_1$ scores of the EfficientNet classifier for the PDR class are also improved with the augmented datasets compared to the original dataset. However, there is no significant difference in $F_1$ and AUC scores of the EfficientNet classifier for augmented datasets with synthetic PDR images of different epochs as indicated in Table \ref{tab:F1 score on training vs test dataset}.

\FloatBarrier
\begin{table}[htp!]
\centering
\caption{\textbf{Assessment of synthetically generated PDR images using the classifiers' $F_1$ scores and AUC scores in augmenting the original imbalanced dataset. $F_1$ scores are recorded for all DR classes.}}
\begin{tabular}{p{1.8cm}p{0.9cm}p{0.5cm}p{0.9cm}p{0.9cm}p{0.9cm}p{0.9cm}p{0.8cm}p{0.7cm}p{1cm}p{1.3cm}}
\toprule
\textbf{Classifier}
    & \textbf{k-fold}
    & \textbf{CW} 
      & \textbf{No DR} 
    & \textbf{Mild NPDR}
    & \textbf{Mod. NPDR} 
    & \textbf{Severe NPDR} 
    & \textbf{PDR} 
    & \textbf{AUC}
    & \textbf{Tr. Tm. (Min.)}
    & \textbf{No. Image Batches}\\
    \midrule
CNN$_{Ref.}$ \cite{Gayathri_2020} 
    & 10-fold 
    & N/A
    & 0.999 
    & 0.999 
    & 1 
    & 0.974 
    & 0.981 
    & N/A
    & N/A 
    & 548 \\
CNN$_{Reimp.}$ 
    & 10 fold 
    & yes 
    & 0.744 
    & 0.001 
    & 0.0003 
    & 0 
    & 0 
    & 0.542
    & 60 
    & 1 \\
CNN$_{Ep.500}$ 
    & 10 fold 
    & yes 
    & 0.744 
    & 0 
    & 0 
    & 0 
    & 0 
    & 0.549
    & 61
    & 1 \\
Effi. Net 
    & N/A 
    & yes 
    & 0.690 
    & 0.156 
    & 0.338 
    & 0.290 
    & 0.395 
    & 0.760
    & 9 
    & 1097 \\
Effi. Net$_{Ep.200}$ 
    & N/A 
    & yes 
    & 0.598 
    & 0.158 
    & 0.336
    & 0.302 
    & 0.408 
    & 0.764
    & 9
    & 1119 \\
Effi. Net$_{Ep.250}$ 
    & N/A 
    & yes
    & 0.643 
    & 0.159 
    & 0.318 
    & 0.281 
    & 0.403 
    & 0.760
    & 9
    & 1119 \\
Effi. Net$_{Ep.500}$ 
    & N/A 
    & yes
    & 0.625 
    & 0.159 
    & 0.303 
    & 0.290 
    & 0.407 
    & 0.760
    & 9
    & 1119 \\
\bottomrule
\multicolumn{11}{l}{Effi Net: EfficientNet; CW: Class Weights; Ep: Epoch number to generate synthetic images for} \\
\multicolumn{11}{l}{augmenting dataset; Mod. NPDR: moderate NPDR; Min: minutes; Ref; Reference Work} \\
\multicolumn{11}{l}{Reimp: Reimplemented for this work; Tr Tm: Training Time} \\
\end{tabular}
\label{tab:F1 score on training vs test dataset}
\end{table}
\FloatBarrier

\section{Conclusion}

This work contributes an empirical interpretation to the selection of synthetic PDR imagery for data augmentation. The contribution of this work is three-fold. First, the selection of suitable evaluation metrics for assessing the similarity and correlation of DR images, representative of their classes and alternate classes. This enabled an effective correlation analysis of PDR images compared to alternate DR images. Second, the selection of suitable evaluation metrics, indicative of their capacity to assess the quality and diversity of DCGAN-based synthetic PDR images and their correlation with classifier performance, is critically assessed. This enabled a quantitative selection of synthetic imagery and an informed augmentation strategy. Third, the selection of synthetic imagery based on the best quality and diversity scores. The efficacy of synthetic images is also evaluated by using them to augment the imbalanced dataset and improve the classification performance of classifiers.

The results demonstrate that MS-SSIM and FID are better at assessing if synthetic imagery belongs to the correct class. The quality of synthetic images is assessed by the FID scores, while diversity is assessed by the MS-SSIM and CD scores. The results indicate the efficacy of synthetic images to augment the imbalanced dataset and improve the $F_1$ score for the PDR class and the AUC score of the EfficientNet classifier. 

This work concludes that evaluation metrics such as MS-SSIM, CD, and FID have a significant impact on assessing the quality and diversity of synthetic images in the biomedical imagery domain. It is important to analyze the impact of different image resolutions, more training epochs, and the lower and upper bound of these metric values for synthetic biomedical imagery, which will be explored as part of future work.       


\end{document}